\documentclass[
reprint,
superscriptaddress,
amsmath,amssymb,
prl
]{revtex4-2}

\usepackage{graphicx}
\usepackage{dcolumn}
\usepackage{bm}
\usepackage[bb=dsserif]{mathalpha} 
\usepackage{hyperref}
\usepackage{mlmodern}
\usepackage[dvipsnames]{xcolor}

\graphicspath{{Figures/}}

\newcommand{\dd}{\mathrm{d}}
\newcommand{\HH}{\hat{\mathcal{H}}}
\newcommand{\mop}{\hat{s}}
\newcommand{\mf}{\omega_\text{m}}
\newcommand{\sq}{r}
\newcommand{\md}{\Delta_\text{m}}

\newcommand{\cd}{\Delta_\text{c}}
\newcommand{\gmc}{g_\text{mc}}

\newcommand{\mdSWT}{\tilde{\Delta}_\text{m}}
\newcommand{\sqSWT}{\tilde{r}}
\newcommand{\gSWT}{g_{12}}
\newcommand{\GSWT}{G_{12}}
\newcommand{\mophB}{\hat{B}}
\newcommand{\Bd}{\Delta_\text{B}}

\begin{document}

\title{\textbf{Quantum State Preparation of Ferromagnetic Magnons by Parametric Driving} 
}

\author{Monika E. Mycroft}
 \affiliation{Institute for Theoretical Physics, Technische Universit\"at Wien, 1040 Vienna, Austria}
 
\author{Rostyslav O. Serha} 
\affiliation{Faculty of Physics, University of Vienna, 1090 Vienna, Austria}

\author{Andrii V. Chumak}
\affiliation{Faculty of Physics, University of Vienna, 1090 Vienna, Austria}

\author{Carlos Gonzalez-Ballestero}%
 \email{Contact author: carlos.gonzalez-ballestero@tuwien.ac.at}
\affiliation{Institute for Theoretical Physics, Technische Universit\"at Wien, 1040 Vienna, Austria}

\date{\today}

\begin{abstract}
We propose a method to prepare and certify Gaussian quantum states of the ferromagnetic resonance spin-wave modes in ferromagnets using a longitudinal drive. Contrary to quantum optics-based strategies, our approach harnesses a purely magnonic feature---the spin-wave nonlinearity---to generate magnon squeezing. This resource is used to prepare vacuum-squeezed states, as well as entangled states between modes of different magnets coupled via a microwave cavity. We propose methods to detect such states with classical methods, such as ferromagnetic resonance or local pickup coils, and quantify the required detection efficiency. We analytically solve the case of ellipsoidal yttrium iron garnet ferrimagnets, but our method applies to a vast range of shapes and sizes.
Our work enables quantum magnonics experiments without single-magnon sources or detectors (qubits), thus bringing the quantum regime within reach of the wider magnonics community. 
\end{abstract}

\maketitle

Quantum magnonics aims to prepare and certify quantum states of the magnetization modes---spin waves, or magnons---in magnetically ordered media~\cite{Lachance-Quirion_2019, YUAN20221, ZARERAMESHTI20221}. Beyond potential applications in hybrid quantum science~\cite{PhysRevB.98.241406, Lachance-Quirion_2019, PhysRevB.101.125404, PhysRevLett.124.093602, PhysRevX.11.031053, PRXQuantum.2.040344, PhysRevLett.128.183603, wu2025microwave, YUAN20221, bruhlmann2025classical},  collective quantum states involving a macroscopic  number of spins ($\sim10^{20}$) could offer fundamental insight into the thermodynamic limit of spin systems or the quantum-classical boundary~\cite{klimov2015quantum, zhang2019quantum, bamba2022magnonic, xu2024macroscopic}.
So far, most proposals for preparing quantum states directly import techniques from quantum optics, mainly the coupling of magnons---usually in yttrium iron garnet (YIG) due to its low damping---to microwave-frequency qubits to prepare few-magnon states~\cite{flebus2018quantum, Lachance-Quirion_2019, fukami2021opportunities, li2021quantum, gonzalez2022towards, YUAN20221, kounalakis2022analog, sharma2022protocol, ZARERAMESHTI20221, zou2022bell, li2023quantum, fukami2024magnon, wu2025microwave, dols2024magnon, li2025solid, dey2025coupling, williams2025generating, dols2025steady, xue2025directional, liu2025magnon}. This has been the chosen route in recent experiments showing quantum behavior of magnons in mm-sized YIG spheres~\cite{lachance2020entanglement, xu2023quantum, xu2024macroscopic}. Due to the challenges of interfacing magnons with microwave qubits, which are sensitive to magnetic noise, these experiments are involved and remain within reach of only a few labs worldwide. 

\begin{figure}[t!]
	\includegraphics[width=\linewidth]{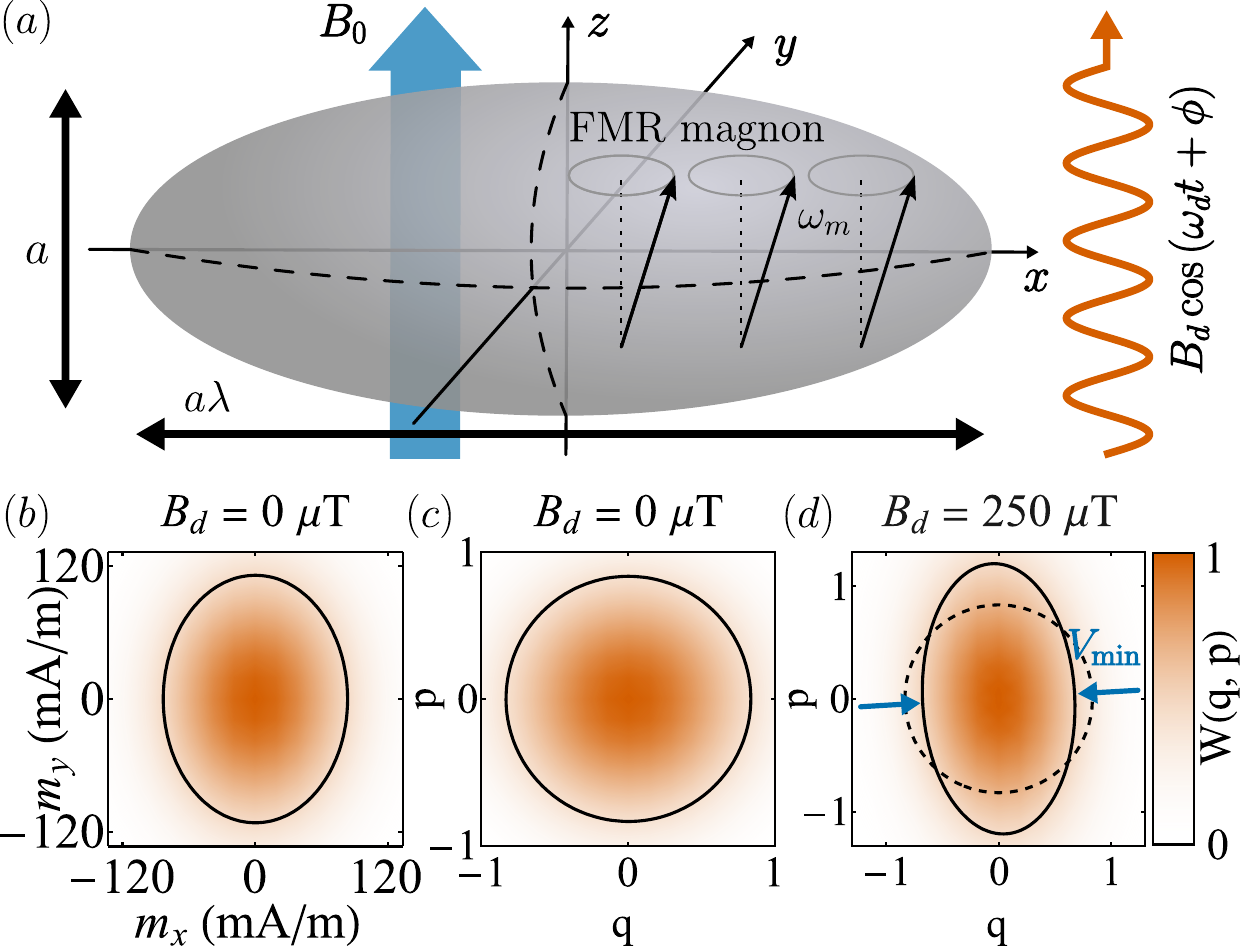}
	\caption{\label{fig:1}(a) The ferromagnetic resonance (FMR)  mode of a ferromagnetic ellipsoid saturated by an homogeneous magnetic field $B_0$ along a short axis can be prepared in a vacuum-squeezed state by a parametric drive applied along the same axis. (b-d) FMR mode Wigner function at $T<400~\text{mK}$. In the absence of parametric drive the magnetization vacuum fluctuations are anisotropic (b) but the state is isotropic in quadrature representation (c), thus showing no quantum squeezing. Under parametric drive with frequency $\omega_\text{d} \approx 2\omega_\text{m}$ the state becomes squeezed, as fluctuations of certain quadratures decrease below the ground-state value (d). Black lines in (b-d) show the Wigner function at half maximum.}
\end{figure}

There is, however, a family of quantum states---Gaussian states~\cite{serafini2023quantum}---which can be prepared using only linear dynamics and does not require qubits. A paramount example are vacuum-squeezed states, for which the variance of a certain quadrature (for spin waves, a given component of the magnetization field) is lower than its ground-state value (Fig.~\ref{fig:1}(d)). Vacuum-squeezed states and their two-mode generalization, Gaussian entangled states, are certifiably nonclassical, as their Glauber-Sudarshan quasiprobability distribution shows negative values
~\cite{WallsMilburn2008}. This allows their use as building blocks for continuous-variable quantum protocols~\cite{braunstein2005quantum}, quantum-enhanced sensing~\cite{caves1981quantum, unruh1983quantum, jia2024squeezing}, or in Bell tests of hidden-variable theories~\cite{kurochkin2014distillation}, as suggested by seminal works~\cite{elyasi2020resources, sharma2021spin, xie2023stationary}. 
Enabling the preparation and detection of spin-wave Gaussian states would remove the need for qubits, allowing many current classical setups to reach the quantum regime. Moreover, Gaussian states would be especially suited for these setups as, for a fixed energy and decoherence budget, they saturate the maximum amount of squeezing and entanglement that can be generated and detected using linear measurements~\cite{WolfPRL2006}.

In this Letter, we propose a method to prepare and measure Gaussian vacuum-squeezed and entangled states of the ferromagnetic resonance (FMR) mode of millimeter-sized magnets using only external magnetic fields, specifically a parametric drive, and linear classical measurements. 
Being realizable on most current experimental platforms, our work enables exploration of core open questions in quantum magnonics, such as what the fundamental limits to quantum state certification or the nature of magnon quantum decoherence are.

We consider a prolate ellipsoidal magnet with volume $V$ centered at the origin, with semiaxes $(a,a,a\lambda)$ ($\lambda>1$) and with the long axis lying along the $x$ axis (Fig.~\ref{fig:1}a).
A uniform magnetic field $B_0$ along $z$ saturates the magnetization to its saturation value $M_\text{S}$.
The magnetization field in the spin-wave approximation 
reads $\mathbf{M}(\mathbf{r},t)\approx M_\text{S}\mathbf{e}_z+\mathbf{m}(\mathbf{r},t)$ where $\mathbf{m}(\mathbf{r},t)$ fulfills $\mathbf{m}(\mathbf{r},t)\cdot\mathbf{e}_z=0$ and $\vert \mathbf{m}(\mathbf{r},t)\vert \ll M_\text{S}$ and obeys the linearized Landau-Lifshitz equation~\cite{stancil2009spin, landau2013course}
\begin{equation}\label{eq:LL}
	\dd\mathbf{m}(\mathbf{r}, t)/\dd t = \omega_0 \mathbf{e_z} \times \mathbf{m}(\mathbf{r}, t) - \omega_\text{M} \mathbf{e_z} \times \mathbf{h}(\mathbf{r}, t).
\end{equation}
Here $\omega_\text{M} \equiv \vert\gamma\vert\mu_0 M_\text{S}$ and $\omega_0 \equiv \vert\gamma\vert B_0 - \omega_\text{M}D_y$ with $\mu_0$ the vacuum permeability and $\gamma$ the gyromagnetic ratio. The field $\mathbf{h}(\mathbf{r}, t)$ is the dipolar magnetic field generated by $\mathbf{m}(\mathbf{r}, t)$, and $D_x$ and $D_y$ are the demagnetizing factors along the $x$ and $y$ axes, with
$D_x  =1-2D_y= \left(\lambda^2-1\right)^{-1}\left(\zeta^{-1}\text{arctanh}(\zeta) - 1\right)$ and $\zeta \equiv \left(\lambda^2 - 1\right)^{1/2}/\lambda$. In Eq.~\eqref{eq:LL} we have used the magnetostatic approximation and neglected magnetocrystalline anisotropy and exchange interaction, which is valid for $\sim$mm-sized samples~\cite{aharoni2000introduction, stancil2009spin}. We have also neglected losses (Gilbert damping) which, as usual in quantum magnonics~\cite{fukami2021opportunities, gonzalez2022towards, kounalakis2022analog, sharma2022protocol, li2025solid}, will be included after quantization.

We first describe the quantum magnon dynamics  in the absence of drive. Since throughout this work the magnon interacts only with fields with wavelength $\gg a\lambda$ and---potentially---detectors in the far field, it is sufficient  to consider only the fundamental eigenmode of Eq.~\eqref{eq:LL}, namely the uniformly magnetized FMR mode. The FMR mode is canonically quantized in terms of harmonic oscillator adimensional quadratures, $\hat q$ and $\hat p$ with $\left[\hat q, \hat p\right] = i$~\cite{MILLS200616,PhysRevB.101.125404}, such that the 
Hamiltonian generating Eq.~\eqref{eq:LL}, namely $\HH_0 \equiv \left(\mu_0/2\right)\int \dd V \mathbf{\hat m}(\mathbf{r}, t) \cdot(\mathbf{\hat m}(\mathbf{r}, t)\omega_0/\omega_\text{M} - \mathbf{\hat h}(\mathbf{r}, t))$,
takes the canonical harmonic oscillator form $\HH_0 =\hbar\mf(\hat q^2+\hat p^2)/2$, with
$\mf \equiv \sqrt{(\omega_0+\omega_\text{M} D_y)(\omega_0+\omega_\text{M} D_x)}$
the FMR mode frequency. The Schr\"odinger Picture magnetization  operator reads
\begin{equation}\label{eq:m}
    \hat{\mathbf{m}} = \mathcal{M}_\text{FMR}\sqrt{2}\left(\hat q, -\sqrt{\xi}\hat p, 0\right), 
\end{equation}
where $\xi \equiv \left(\omega_0 + \omega_\text{M}D_x\right)/\left(\omega_0 + \omega_\text{M}D_y\right)<1$ is the magnetization asymmetry parameter ($\xi\to 1$ for a spherical magnet) and $\mathcal{M}_\text{FMR}^2 \equiv \hbar\vert\gamma\vert M_\text{S}/\left(2V\sqrt{\xi}\right)$ the zero-point magnetization, which quantifies the ground-state variance $\langle\hat m_x^2\rangle=\xi\langle \hat m_y^2 \rangle=\mathcal{M}_\text{FMR}^2$. 
In the presence of  damping the magnon quantum state, described by the density operator $\hat \rho(t)$, is governed by the master equation~\cite{PhysRevB.101.125404,gonzalez2022towards}
\begin{equation}\label{eq:master_sq}
     \dot{\hat{\rho}}
     = -i\hbar^{-1}[\HH_0, \hat \rho]  + \gamma_\text{m}\mathcal{L}_{\mop\mop^\dagger}\left[\hat \rho\right].
 \end{equation}
Here, $\mathcal{L}_{\hat a \hat b}\left[\hat \rho\right] \equiv \hat a \hat \rho \hat b - (\hat b\hat a\hat \rho+\hat \rho\hat b\hat a)/2$ is the Lindblad dissipator, $\mop \equiv \left(\hat q + i\hat p\right)/\sqrt{2}$ the magnon ladder operator, and $\gamma_\text{m}$ the Gilbert damping rate.
Eq.~\eqref{eq:master_sq} is valid in the low-temperature limit $T\ll \hbar \mf/k_\text{B} \sim400~\text{mK}$, which is needed for our state preparation method and which we assume hereafter.
From Eq.~\eqref{eq:master_sq} expectation values for all observables can be obtained. 

The steady-state of Eq.~\eqref{eq:master_sq} is the ground state (magnon vacuum), which, like in any quantum harmonic oscillator, has isotropic quadrature variances $\langle\hat q^2\rangle=\langle\hat p^2\rangle=1/2$ (see Fig.~\ref{fig:1}(c)) and does not show non-classical features. 
Note that the vacuum state still shows anisotropic magnetization fluctuations (see Eq.~\eqref{eq:m} and Fig.~\ref{fig:1}(b)) purely due to geometry.
A vacuum-squeezed state is a Gaussian state for which the variance of any quadrature is smaller than the vacuum variance, that is, a state for which the generalized quadrature $\hat X_\theta\equiv \left(e^{i\theta} \mop^\dagger + e^{-i\theta}\mop\right)/\sqrt{2}$ (note that $\hat{X}_{0}=\hat{q}$ and $\hat{X}_{\pi/2}=\hat{p}$) has variance
$\langle\hat X_\theta^2\rangle<1/2$ for some value of $\theta$ [Fig.~\ref{fig:1}(d)]. 
To prepare such state, we propose to apply a homogeneous parametric drive field $\mathbf{B}_\text{d}(t)\equiv B_\text{d}\cos\left(\omega_\text{d}t+\phi\right)\mathbf{e}_z$, changing the magnon Hamiltonian to $\HH \equiv \HH_0+\HH_\text{sq}(t)$, with the Zeeman term $\HH_\text{sq}(t) = - \int\dd V \mathbf{\hat{M}}(\mathbf{r}) \cdot \mathbf{B}_\text{d}(t) \approx -V\hat{M}_zB_\text{d}\cos\left(\omega_\text{d}t+\phi\right)$~\footnote{Note that this approximation neglects potential excitation of spurious, long-lived high-wavevector modes above the parametric instability threshold~\cite{serha2025ultra}. This approximation is valid since we assume the amplitude $B_\text{d}$ to remain always below threshold.}. We simplify this term as follows: (i) we introduce the $z-$component of the FMR mode magnetization up to second order in  $\hat{\mathbf{m}}/M_\text{S}$, i.e., 
 $\hat{M}_z \approx M_\text{S} - \hat{\mathbf{m}}^2/(2M_\text{S})$~\cite{stancil2009spin}. The term $\propto \hat{\mathbf{m}}^2$ stems from the nonlinear nature of spins and is key to generating squeezing. (ii) 
We neglect terms oscillating faster than $\omega_\text{d}$ in the interaction picture (rotating wave approximation), valid provided that $\vert r\vert, \vert 2\omega_\text{m}-\omega_\text{d}\vert \ll \omega_\text{d}$ with $r$ defined below. (iii) We transform to a rotating frame via the unitary transformation $\mathcal{U}(t) =  \exp\left[i\omega_\text{d}t\mop^\dagger\mop/2\right]$. In this frame the Hamiltonian becomes time-independent, 
\begin{equation}\label{eq:Hsq}
			\HH = \hbar \md \mop^\dagger\mop + \hbar \left(re^{i\phi}\mop^2 + \text{H.c.}\right),
		\end{equation}
where $\md \equiv \mf-\omega_\text{d}/2$ and with the squeezing rate $r\equiv \left(\vert\gamma\vert B_\text{d}/8\right)\left(1/\sqrt{\xi}-\sqrt{\xi}\right)$. Note that the anisotropic magnet geometry is crucial as this rate vanishes for $\xi=1$ (e.g. for spheres). Introducing the Hamiltonian Eq.~\eqref{eq:Hsq} into Eq.~\eqref{eq:master_sq} we obtain the minimum variance in the steady state, 
\begin{equation}\label{eq:R}
	V_{\rm min}\equiv\min_\theta\langle\hat X_\theta^2\rangle = \frac{1/2}{1 + 4r\left(\gamma_\text{m}^2 + 4\md^2\right)^{-1/2}} <\frac{1}{2},
\end{equation}
indicating steady-state vacuum squeezing. The minimum variance occurs at a phase-space angle $2\theta_{\rm sq} = -\phi + \arctan\left(\gamma_\text{m}/\left(2\md\right)\right)$, which is tunable through the detuning $\md$ and phase $\phi$ of the parametric drive, allowing the preparation of arbitrary vacuum-squeezed states.

\begin{figure}[t]
	\includegraphics[width=\linewidth]{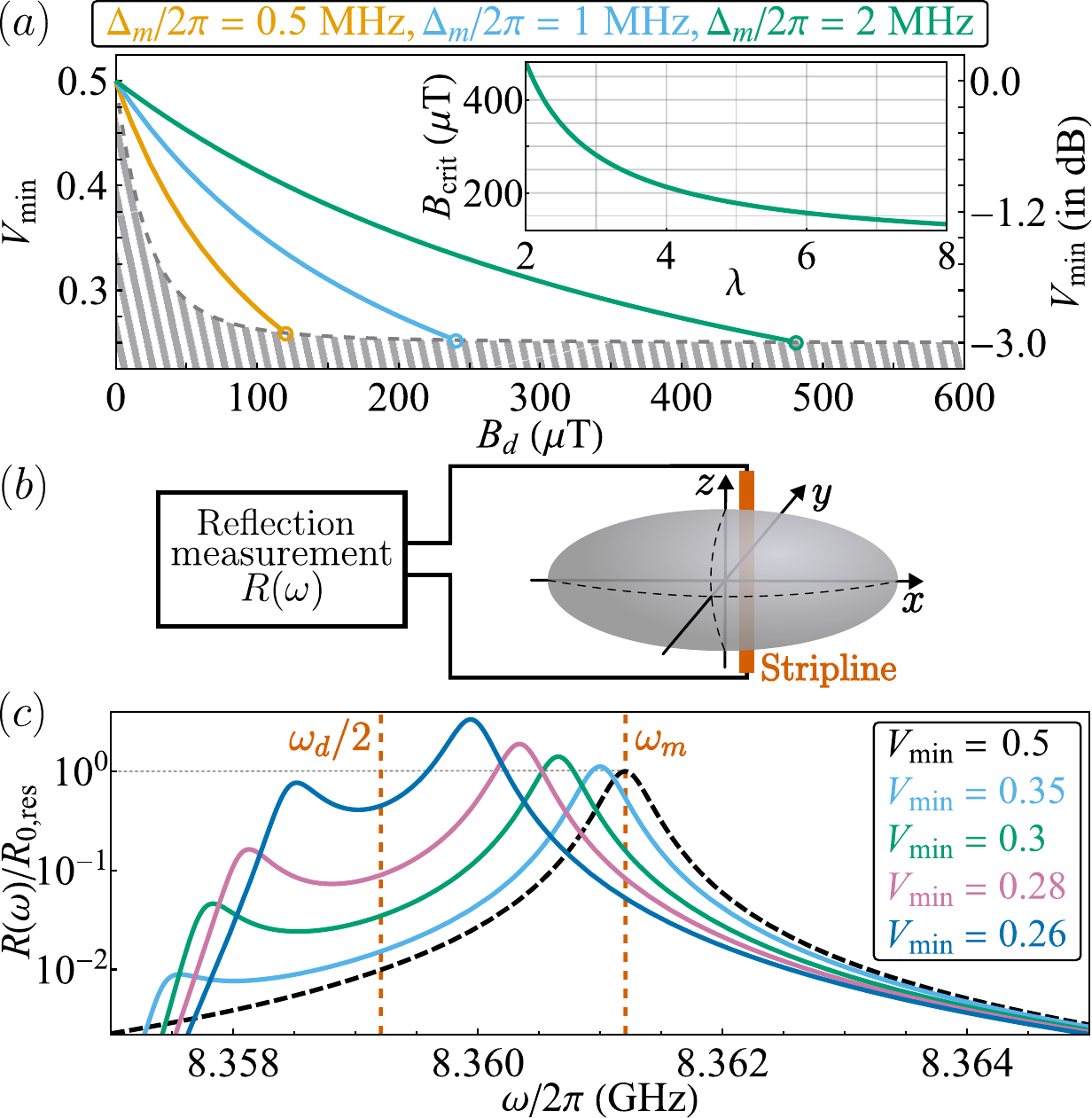}
	\caption{\label{fig:2}(a) Squeezing measure, Eq.~\eqref{eq:R}, versus parametric drive strength for various detunings $\md$ and $\phi=0$. Hatched area marks parametric instability. Inset: parametric instability threshold amplitude $B_\text{d}$ versus ellipsoid aspect ratio. (b) Vacuum squeezing can be detected by microwave reflection. (c) Reflectivity spectrum for $\md/2\pi = 2~\text{MHz}$ and different degrees of squeezing (dashed line corresponds to no squeezing). In this and other figures we take  the following parameters for YIG: $\gamma = -1.76 \times 10^{11}~\text{T}^{-1}\text{s}^{-1}  $, 
    $M_\text{S} = 5870~\text{kA}/\text{m}$, $\gamma_\text{m}/2\pi = 0.4~\text{MHz}$, $B_0 = 0.4~\text{T}$, and $\lambda=2$.}
\end{figure}

Fig.~\ref{fig:2}(a) shows the minimum variance $V_{\rm min}$ as a function of parametric drive amplitude $B_\text{d}$. The hatched area marks the regime $\left\vert 2r\right\vert > \left\vert\md\right\vert$ where the Hamiltonian Eq.~\eqref{eq:Hsq} becomes dynamically unstable~\cite{kustura2019quadratic} and the steady state of Eq.~\eqref{eq:master_sq} does not exist. This is the well-known magnon parametric instability~\cite{Suhl1957, Zakharov1975, chumak2009parametrically}. Within the stability regime, the maximum achievable squeezing, attained for $\vert\md\vert\gg \gamma_\text{m}$, corresponds to $V_{\rm min} = \left(1/2\right)(1+2 r/\vert \md\vert)^{-1}$. Near the instability boundary, it reaches the well-known 3~dB  bound for parametric squeezing~\cite{gardiner2004quantum}  ($V_{\rm min}\to 1/4$). This boundary---and thus maximal squeezing---is reachable at every detuning $\md\gtrsim \gamma_\text{m}$, but larger detunings require larger squeezing rates $r$, i.e., larger drive amplitudes $B_\text{d}$ or aspect ratios $\lambda$ (see Fig. \ref{fig:2}(b)). 
We propose to detect magnon squeezing via a conventional FMR resonance experiment (Fig. \ref{fig:2}(b)), i.e., through its impact on the microwave reflection spectrum. The reflectivity is computed using input-output formalism (see End Matter) and reads
\begin{equation}\label{eq:reflectivity}
    R(\omega) = R_{0,\rm res}\frac{\gamma_\text{m}^2}{4}\frac{\gamma_\text{m}^2/4 + \left(\Delta_\text{p} + \md\right)^2}{\Delta_\text{p}^2\gamma_\text{m}^2 + \left(\Bd^2 - \Delta_\text{p}^2 + \gamma_\text{m}^2/4\right)^2},
\end{equation}
where we define $\Delta_\text{p}\equiv \omega-\omega_\text{d}/2$ and $\Bd^2 \equiv\md^2-4r^2$, and where $R_{0,\rm res}$ is the peak-value reflectivity in the absence of squeezing (i.e. $R(\omega_\text{m})$ at $r\to 0$).
As shown in Fig. \ref{fig:2}(c), the characteristic Lorentzian shape of the FMR reflectivity (dashed black line) evolves toward a two-peaked structure centered at $\omega=\omega_\text{d}/2$ as the degree of magnon squeezing increases. This indicates the onset of microwave parametric amplification~\cite{collett1984squeezing}, a core signature of squeezing that has been experimentally observed e.g.\ in photonic systems~\cite{WallsMilburn2008}.

\begin{figure}[t]
	\includegraphics[width=\linewidth]{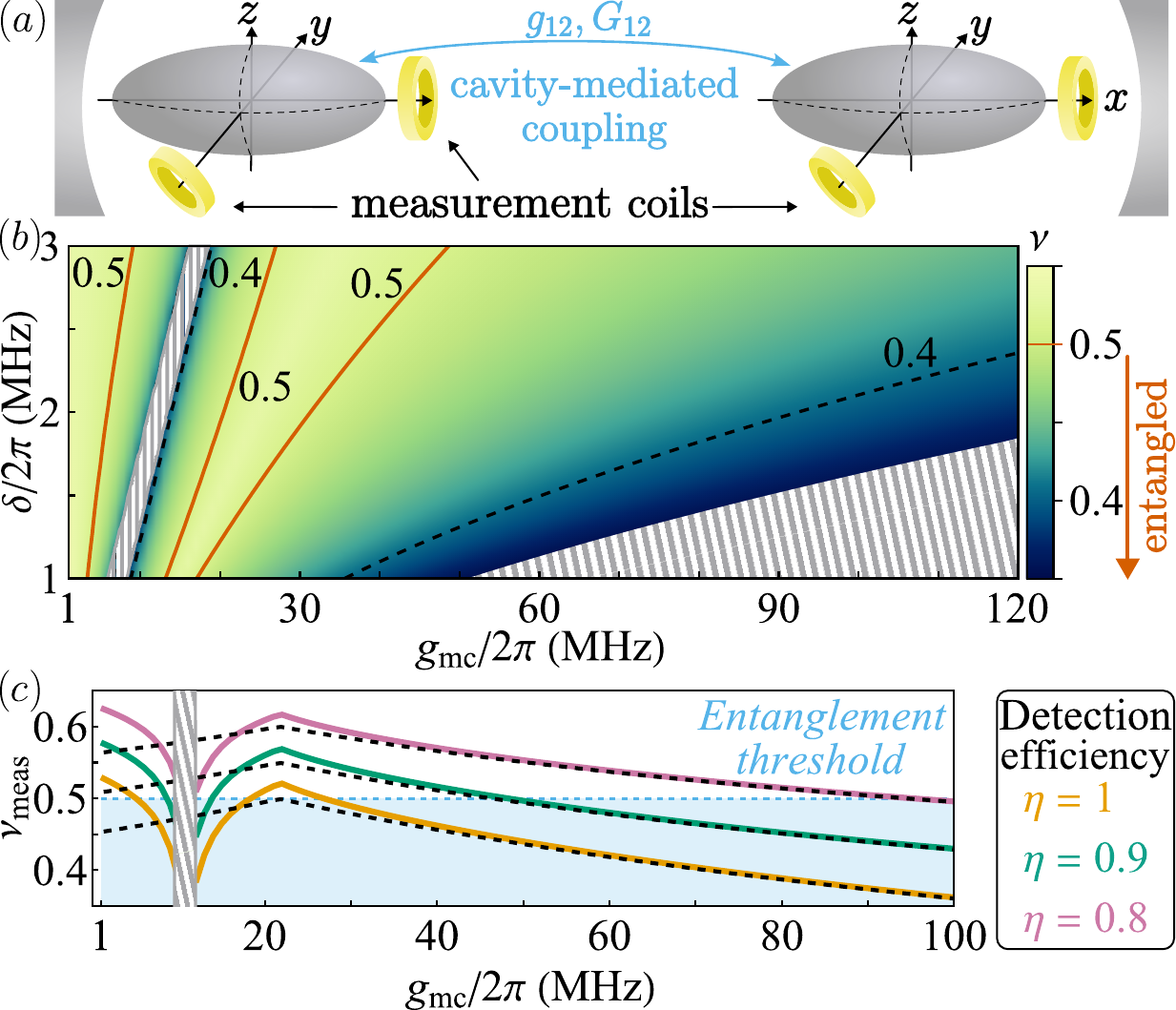}
	\caption{\label{fig:3}(a) Scheme for preparing entangled magnon states and detecting them using pickup coils.  
    (b) Steady-state entanglement versus magnon-cavity coupling rate $g_{\rm mc}$ and drive detuning from the instability threshold $\delta$, for $\md/2\pi = 2~\text{MHz}$, $\cd = \md + 10 \gmc$ and $\kappa = 10^{-4}\omega_\text{c}$. Hatched areas indicate collective instability regions. (c) Measured entanglement for $\delta/2\pi = 1.5~\text{MHz}$ and for different coil detection efficiencies. Dashed curves show the analytic approximation Eq.~\eqref{eq:nu_eta}. }
\end{figure}

Vacuum squeezing can also be used as a resource to prepare Gaussian entangled states between oscillators coupled through particle-conserving interactions~\cite{scheel2001entanglement, kim2002entanglement,xiang2002theorem}. Let us consider two identical magnetic ellipsoids subject to identical static and parametric drive fields $B_0\mathbf{e}_z$ and $B_\text{d}\cos(\omega_\text{d} t)\mathbf{e}_z$. Both are placed within a single-mode microwave cavity with frequency $\omega_\text{c}$ and decay rate $\kappa$, which couples to their FMR modes at a rate $g_{\rm mc}$ and thus mediates the interaction between them (Fig.~\ref{fig:3}(a)).   
In the frame rotating at frequency $\omega_\text{d}/2$, the master equation for the density operator $\hat \rho_{\rm mc}$ of the three degrees of freedom (two magnon modes and a cavity mode) 
reads $\dd\hat \rho_{\rm mc}/\dd t = -(i/\hbar)\left[\HH_1+\HH_2+\HH_c+\HH_{\rm int},\hat \rho_{\rm mc}\right]+\gamma_\text{m}\sum_{j=1,2}\mathcal{L}_{\mop_j\mop^\dagger_j}[\hat \rho_{\rm mc}]+\kappa \mathcal{L}_{\hat c\hat c^\dagger}[\hat \rho_{\rm mc}]$, with $\mop_j$ and $\hat c$ the ladder operators of magnon $j$ and cavity mode, respectively. The total Hamiltonian  contains the magnon Hamiltonians $\HH_j$, given by Eq.~\eqref{eq:Hsq}, the free cavity Hamiltonian
$\HH_c = \hbar\cd\hat{c}^\dagger\hat{c}$ with $\cd\equiv\omega_\text{c}-\omega_\text{d}/2$, and the magnon-cavity interaction $\HH_\text{int} = \hbar \gmc\left(\hat{c}^\dagger\mop_1 + \hat{c}^\dagger\mop_2\right) + \text{H.c.}$ in the rotating wave approximation, which is valid in the regime $\vert g_{\rm mc}\vert^2 \ll \vert 2\Bd \left(\cd \pm \Bd\right)^2/\left(\md \pm \Bd\right)\vert$. 

To obtain an effective cavity-mediated coupling between the FMR modes of both magnets, we adiabatically eliminate the cavity following four steps (see End Matter): (i) we perform a Bogoliubov transformation to cast the magnon Hamiltonians in diagonal form. This transformation is unitary below the instability threshold ($ \left\vert 2r\right\vert< \left\vert\md\right\vert$ or equivalently $\Bd\in\mathbb{R}$). (ii) The diagonal Hamiltonians allow us to perform a Schrieffer-Wolff unitary transformation~\cite{bravyi2011schrieffer} that decouples the magnons from the cavity up to terms of second order in the perturbative parameter $\left(\gmc/\left(\cd\mp\Bd\right)\right)\left(\left(\md \pm \Bd\right)/\left(2\Bd\right)\right)^{1/2}$, which is assumed small. (iii) we reverse the Bogoliubov transformation [i]. (iv) We undertake the rotating wave approximation to neglect any dissipative coupling between magnons and cavity. The resulting magnon dynamics are decoupled from the cavity, and we write the reduced master equation for the state of the two-magnon subsystem, $\hat \rho_\text{m}$, as
$\dd\hat \rho_\text{m}/\dd t = -(i/\hbar)[\HH_\text{eff},\hat \rho_\text{m}]+\sum_{jk}\Gamma_{jk}\mathcal{L}_{\hat{v}_j\hat{v}_k}[\hat \rho_\text{m}]$, where $\hat{\mathbf{v}}\equiv (\hat s_1, \hat s_1^\dagger, \hat s_2, \hat s_2^\dagger)^T$. The effective Hamiltonian reads
\begin{eqnarray}
    \HH_\text{eff}/\hbar =&& \sum_{j=1,2} \left(\mdSWT\mop_j^\dagger\mop_j+(\sqSWT\mop_j^2 + \text{H.c.})\right) +
    \nonumber\\
    &&+\left( \gSWT \mop_1\mop_2^\dagger + \text{H.c.}\right) + \left(\GSWT\mop_1\mop_2 + \text{H.c.}\right),\label{eq:Heff}
\end{eqnarray}
with rates $\mdSWT \!= \!\gSWT\!\left(\Bd^2 + \md\cd\right)\!/[\Bd\left(\md \!+\! \cd\right)]$, 
$\sqSWT=\allowbreak\left(2\gSWT r/\Bd\right)\!\left[1-\md\cd/\!\left(2\Bd\!\left(\md + \cd\right)\right)\right]$,  and $\gSWT = \allowbreak \GSWT\left(\md+\cd\right)/(2 r)=\allowbreak -\gmc^2\!\left(\md\! + \!\cd\right)\!/\!\left(\cd^2 - \Bd^2\right)$. All these rates depend on cavity parameters and magnon squeezing rates. In the limit $r\to0$ they recover the known cavity magnonics result $\gSWT = -\gmc^2/\left(\cd-\md\right)$ and $\GSWT = 0$~\cite{ZARERAMESHTI20221}. 

To quantify magnon-magnon entanglement, we compute from the master equation the steady-state covariance matrix $\bar{\sigma}_{jk}\equiv \sum_{n,m} \allowbreak \mathbf{T}_{jm}\left(\langle \hat v_m \hat v_n\rangle + \langle \hat v_n \hat v_m\rangle\right) \mathbf{T}_{kn}/2$ with $\mathbf{T}$ the ladder-to-quadrature transformation matrix $\sqrt{2}\mathbf{T} \equiv \mathbb{1}_{2\times2}\otimes \left(\begin{smallmatrix} 1 & 1 \\ -i & i \end{smallmatrix}\right)$. A sufficient and necessary condition for the two modes to be entangled is $\nu < 1/2$, with $\nu$ being the smallest symplectic eigenvalue of the partially transposed covariance matrix~\cite{duan2000inseparability, simon2000peres} (see End Matter for details). We compute $\nu$ numerically and show it in Fig.~\ref{fig:3}(b) as a function of the magnon-cavity coupling $g_{\rm mc}$ and $\delta \equiv \md - 2\sq$. Magnon-magnon entanglement is achievable for a wide range of parameters, and increases with cavity-magnon coupling $g_{\rm mc}$ and at lower values of $\delta$, i.e. when approaching the parametric instability threshold $\delta=0$. 
This evidences that entanglement stems from the combination of (cavity-mediated) magnon-magnon interactions and magnon squeezing. The dashed regions  in Fig.~\ref{fig:3}(b) describe new dynamical instabilities originating either from too strong magnon-magnon coupling~\cite{kustura2019quadratic} (lower right corner) or---in analogy to the already discussed parametric instability---from a strong squeezing of the \textit{collective magnon mode} $\mop_+ \propto \mop_1 + \mop_2$ (thin region, see End Matter for details). In the experimentally relevant regime 
$\vert \GSWT\vert \ll 2\vert\sqSWT\vert$ and  $2\vert\sqSWT\vert, \vert\Gamma_{11}\vert\ll\vert \tilde{\Delta}_m + g_{12}\vert$, and $\vert\Gamma_{11}\cosh(2\alpha_-)\vert^2\ll(\mdSWT-\gSWT)^2-4\sqSWT^2$, and $\vert 4\Gamma_{11}\left(\begin{smallmatrix}
    \tanh{\alpha_-}\\\coth{\alpha_-}
\end{smallmatrix}\right)/\left(\Gamma_{12}-\Gamma_{14}\right)\vert\ll1$, and $2\vert \Gamma_{21}\vert\ll \vert \Gamma_{12}+\Gamma_{14}\vert$
we can analytically approximate (see End Matter)
\begin{equation}\label{eq:nu_0}
	(2\nu)^2 \approx \tilde{\Delta} - \sqrt{\tilde{\Delta}^2 -d},
\end{equation} 
where $\tilde{\Delta} \equiv \left(2n_- + 1\right)\cosh(2\alpha_-) - 2\text{Re}\{z\}\sinh(2\alpha_-)$ and $d = \left(2n_-+1\right)^2 - 4\vert z\vert^2$, with $\tanh\alpha_-=2\tilde{r}/(\tilde{\Delta}_m-g_{12})$, 
$n_-=\sinh^2\alpha_-$, $2z=(\Gamma_{12}-\Gamma_{14})\sinh(2\alpha_-)/(\Gamma_{12}-\Gamma_{14}+4i\vert\tilde{r}\vert(\coth^2\alpha_- - 1)^{1/2})$. 
Eq.~\eqref{eq:nu_0} is an excellent approximation as shown by Fig.~\ref{fig:3}(c) (lowest dashed curve) and thus provides a simple tool to assess the entanglement capabilities of cavity magnonics setups.

A core advantage of magnons (as opposed to e.g.,\ mechanical resonators) is that both quadratures $\hat q$ and $\hat p$ are accessible experimentally as they are proportional to the $x-$ and $y-$ components of the magnetization (see Eq.~\eqref{eq:m}). This enables full-state tomography to detect entanglement using four small pick-up coils threading the corresponding Cartesian axes (Fig.~\ref{fig:3}(a)), each sensing one of the four quadratures $\{\hat q_1,\hat p_1,\hat q_2,\hat p_2\}$ of the two-magnon Gaussian state~\cite{gloppe2019resonant}. The measurement output of each coil is $\hat q_{j, \rm meas} = \sqrt{\eta}\hat{q}_j + \sqrt{1-\eta}\hat{q}_{j,\rm n}$ $(j=1,2)$, and a similar expression for $\hat{p}_{j, \rm meas}$. Here we have introduced the detection efficiency $\eta\in[0,1]$ (related to the low-temperature signal-to-noise ratio $\text{SNR}_0=\eta/(1-\eta)$) and noisy quadratures $\mathbf{\hat{w}}_{\rm n}\equiv\{\hat{q}_{1,\rm n},\hat{p}_{1,\rm n},\hat{q}_{2,\rm n},\hat{p}_{2,\rm n}\}$ modeling detection noise, which are assumed independent, thermal, and uncorrelated from the signal, $\langle \hat{w}_j\hat{w}_k\rangle=\delta_{jk}/2$ and $\langle \hat{v}_j\hat{w}_k\rangle=0$. Experimentally, the two-magnon covariance matrix is estimated by reconstructing the covariance matrix of the measured quadratures, leading to entanglement underestimation due to measurement noise. Indeed, as shown by Fig.~\ref{fig:3}(c) (solid lines), for the chosen parameters, entanglement certification requires $\eta\gtrsim0.8$. This bound is not fundamental and can be improved by parameter optimization. In the validity regime of Eq.~\eqref{eq:nu_0} the measured entanglement can be approximated as (see End Matter)
\begin{equation}\label{eq:nu_eta}
    (2\nu_{\rm meas}(\eta))^2 = (2-\eta)\left(1-\eta+2\eta\nu(0)^2\right),
\end{equation}
where $\nu(0)$ is the true magnon-magnon entanglement Eq.~\eqref{eq:nu_0}. Eq.~\eqref{eq:nu_eta} is shown as dashed lines in  Fig.~\ref{fig:3}(c).

In conclusion, we propose a method to prepare single- and two-mode Gaussian quantum states of FMR modes in magnets. The approach applies to a broad range of magnet sizes and to any geometry with unequal 
$x-$ and $y-$ FMR mode amplitudes, such as in-plane magnetized disks or slabs. Combined with low temperatures and conventional microwave driving, it enables the preparation of arbitrary multimode Gaussian states with up to 3~dB of squeezing. We also introduce certification methods based on standard classical techniques available in most magnonics laboratories. This work opens access to the quantum regime of magnons using conventional setups without qubits. Future work will explore protocols exceeding the 3~dB limit and exploit highly sensitive squeezed states to probe magnon decoherence, which remains poorly understood despite recent experimental interest~\cite{serha2025ultra, serha2025ysgag, serha2025magnetic}. Notably, the squeezed states considered here involve 
$\sim 10^{20}$ spins, vastly exceeding atom-based spin-squeezed systems~\cite{SqueezingSpins}, and offer new opportunities for studying macroscopic entanglement and quantum sensing.

\begin{acknowledgments}
The authors were supported in whole or in part by the Austrian Science Fund (FWF) under Project PAT-1177623 (C.G.B. and M.E.M) and Project No. 10.55776/I6568 (R. O. S. and A. V. C.). We thank J. del Pino, L. Papa, J. J. Garcia-Ripoll, and G. Vitagliano for insightful discussions. 
\end{acknowledgments}

\bibliography{manuscript}
\clearpage
\appendix
\section{Detection of squeezing with FMR}

We consider a magnon mode $\mop$ coupled to three independent boson baths: $\hat b(\omega)$, responsible for intrinsic Gilbert damping at rate $\gamma_\text{m}$, and $\hat a_{1,2}(\omega)$, describing microwave modes propagating in opposite directions along the transmission line and adding to an additional radiative damping rate $\kappa_\text{m} \ll \gamma_\text{m}$. The total Hamiltonian is
\begin{eqnarray}
	&&\HH_\text{tot}/\hbar = \mf\mop^\dagger\mop + \left(\sq e^{i\omega_\text{d}t}\mop^2 + \text{H.c.}\right) + \nonumber\\
    &&+	\int\dd\omega\omega\hat b^\dagger(\omega)\hat b(\omega) - i\sqrt{\frac{\gamma_\text{m}}{2\pi}}\int\dd\omega\left(\mop^\dagger\hat b(\omega) - \text{H.c}\right) +\\
	&&+ \sum_{j=1,2}\!\left(\int\!\dd\omega\omega\hat a_j^\dagger(\omega)\hat a_j(\omega) - i\sqrt{\frac{\kappa_\text{m}}{4\pi}}\int\!\dd\omega\left(\mop^\dagger\hat a_j(\omega) - \text{H.c.}\right)\right). \nonumber
\end{eqnarray}
Transforming to a rotating frame $\mop'(t) = \mop(t) e^{i\omega_\text{d}t/2}$ and defining $\mathbf{\mop'}(t) \equiv \left(\mop(t), \mop^\dagger(t)\right)^T$ (and similarly for the other operators), we obtain the Heisenberg equation of motion  
\begin{eqnarray}
    \frac{\dd}{\dd t}\mathbf{\mop'} =&& \mathbf{M}\mop'(t) - \sqrt{\gamma_\text{m}}\mathbf{\hat b_\text{in}'}(t) - \sqrt{\frac{\kappa_\text{m}}{2}}\!\sum_{j=1,2}\!\mathbf{\hat a}_{\text{in},j}'(t),
\end{eqnarray}
where 
\begin{equation}
    \mathbf{M} \equiv \begin{pmatrix}
        	-i\md - \Gamma/2 & - 2 i \sq \\
			2 i \sq & i\md - \Gamma/2
    \end{pmatrix},
\end{equation}
and where the input modes are defined as~\cite{gardiner2004quantum} $\hat b_\text{in}(t) \equiv (2\pi)^{-1/2}\int \dd \omega \hat b(\omega,t)e^{-i\omega(t-t_0)}$ for an initial time $t_0<t$ in the distant past (and similarly for $\hat a_{\text{in},j}$). $\Gamma\equiv\gamma_\text{m} + \kappa_\text{m}$ is the total magnon dissipation rate.
In the frequency domain, the magnon operator obeys
\begin{eqnarray}\label{eq:internal_mode}
    \mop(\omega_\text{d}/2 + \omega) =&& \mathbf{e_1}^T\left(\mathbf{M} + i\omega\right)^{-1}\left(\sqrt{\gamma_\text{m}}\mathbf{\hat b_\text{in}'}(\omega)\right. +\nonumber\\
    &&+\left.\sqrt{\kappa_\text{m}/2}\left(\mathbf{\hat a_{\text{in},1}'}(\omega) + \mathbf{\hat a_{\text{in},2}'}(\omega)\right)\right)
\end{eqnarray}
where $\mathbf{\mop'}(\omega) \equiv (2\pi)^{-1/2}\int\dd t \mathbf{\mop'}(t) e^{i\omega t} = (\mop(\omega_\text{d}/2 + \omega),   \mop^\dagger(\omega_\text{d}/2 - \omega))^T$ and similarly for the input operators, and the two-dimensional unit vector $\mathbf{e}_j^T\equiv (\delta_{j1},\delta_{j2})$. 

We use Eq.~\eqref{eq:internal_mode} and its distant-future analog to derive the input-output relation~\cite{gardiner2004quantum}, $\hat a_{\text{out},2}(\omega) = \sqrt{\kappa_\text{m}/2}\mop(\omega) + \hat a_{\text{in},2}(\omega)$ with $\hat a_{\text{out},2}(t) \equiv (2\pi)^{-1/2}\int \dd \omega \hat a_2(\omega,t)e^{-i\omega(t-t_1)}$ for some distant future time $t_1>t$ (and similarly for the other output modes). Substituting Eq.~\eqref{eq:internal_mode} explicitly yields $\hat a_{\text{out},2}(\omega_\text{d}/2 + \omega) = \mathbf{\Lambda}^T(\omega)\cdot\mathbf{\hat u}_\text{in}(\omega)$ where $\mathbf{\hat u}_\text{in}(\omega) \equiv (\hat b_\text{in}(\omega_\text{d}/2 + \omega),\hat b_\text{in}^\dagger(\omega_\text{d}/2 - \omega), \hat a_{\text{in},1}(\omega_\text{d}/2 + \omega),\hat a_{\text{in},1}^\dagger(\omega_\text{d}/2 - \omega),\hat a_{\text{in},2}(\omega_\text{d}/2 + \omega),\hat a_{\text{in},2}^\dagger(\omega_\text{d}/2 - \omega))^T$,
and $\mathbf{\Lambda}(\omega) \equiv (\sqrt{\kappa_\text{m}\gamma_\text{m}/2} \mathbf{P}_{11}(\omega), \sqrt{\kappa_\text{m}\gamma_\text{m}/2} \mathbf{P}_{12}(\omega), \left(\kappa_\text{m}/2\right)\mathbf{P}_{11}(\omega),\\ \left(\kappa_\text{m}/2\right)\mathbf{P}_{12}(\omega), 1 + \left(\kappa_\text{m}/2\right)\mathbf{P}_{11}(\omega), 0)^T$ with $\mathbf{P}_{ij}(\omega) \equiv \mathbf{e}_i^T\left(\mathbf{M} + i\omega\mathbb{1}\right)^{-1}\mathbf{e}_j$. The  two-frequency correlation function of the output field can then be written as
\begin{eqnarray}\label{eq:output}
    \langle\hat a_{\text{out},2}^\dagger(\omega_\text{d}/2 &&+ \omega) \hat a_{\text{out},2}(\omega_\text{d}/2 + \omega')\rangle =\nonumber\\&&= \sum_{\mu, \nu}\mathbf{\Lambda}^\ast_\mu(\omega)\mathbf{\Lambda}_\nu(\omega')\langle \mathbf{\hat u}_\mu^\dagger(\omega)\mathbf{\hat u}_\nu(\omega')\rangle.
\end{eqnarray}
In an FMR configuration, we assume that all baths are in thermal equilibrium and that the bath $\hat a_{\text{in},1}(\omega)$ is driven by a classical probe with amplitude $\alpha_\text{p}$ at frequency $\omega_\text{p}$. In this case, the only nonvanishing correlation functions on the right-hand side of Eq.~\eqref{eq:output} are
\begin{eqnarray}
	\langle \hat b_{\text{in}}^\dagger(\omega)\hat b_{\text{in}}(\omega')\rangle &&= \bar n(\omega)\delta(\omega-\omega')\\
    \langle b_{\text{in}}(\omega)\hat b_{\text{in}}^\dagger(\omega')\rangle &&= \langle \hat b_{\text{in}}^\dagger(\omega)\hat b_{\text{in}}(\omega')\rangle + \delta(\omega-\omega')
\end{eqnarray}
and equivalent terms for the input mode $\hat a_{\text{in},2}(\omega)$, and
\begin{multline}\label{eq:input}
    \langle \hat a_{\text{in},1}^\dagger(\omega)\hat a_{\text{in},1}(\omega')\rangle =\\=\left(\bar n(\omega) + \vert\alpha_\text{p}\vert^2 \delta(\omega_\text{p}-\omega)\right)\delta(\omega-\omega')
\end{multline}
together with $\langle \hat a_{\text{in},1}(\omega)\hat a_{\text{in},1}(\omega')\rangle = \alpha_\text{p}^2 \delta(\omega_\text{p} - \omega)\delta(\omega+\omega')$ and their complex conjugates. Here, $\bar n(\omega) \equiv \left(\exp\left[\hbar\omega/k_\text{B}T\right] - 1\right)^{-1}$ is the Bose-Einstein distribution.

The measured FMR reflection is obtained from 
\begin{equation}
    R(\omega_\text{p}) = \lim_{\epsilon\to 0} \frac{\langle\hat a_{\text{out},2}^\dagger(\omega_\text{p})\hat a_{\text{out},2}(\omega_\text{p} + \epsilon)\rangle}{\langle\hat a_{\text{in},1}^\dagger(\omega_\text{p})\hat a_{\text{in},1}(\omega_\text{p} + \epsilon)\rangle}.
\end{equation}
Using Eq.~\eqref{eq:output} and \eqref{eq:input}, we obtain (for $\omega_\text{p} \neq \omega_\text{d}/2$)
\begin{equation}
		R(\omega_\text{p}) = \frac{\kappa_\text{m}^2}{4}\big| \mathbf{P}_{11}(\omega_\text{p}-\omega_\text{d}/2)\big|^2.
	\end{equation}
Eq.~\eqref{eq:reflectivity} is obtained by taking the physical limit $\gamma_\text{m} \gg \kappa_\text{m}$ and defining $R_{0,\text{res}} \equiv R(\omega_\text{m})\vert_{r=0}=\kappa_\text{m}^2/\gamma_\text{m}^2$. 
	
\section{Derivation of the two-magnon master equation}
We apply the Bogoliubov transformation $\mop_{\text{B},j} = c_\alpha \mop_j + s_\alpha\mop_j^\dagger$, with $c_\alpha\equiv \cosh\alpha$ and $s_\alpha\equiv \sinh\alpha$,
to the total Hamiltonian $\HH_\text{tot} = \HH_1 + \HH_2 + \HH_c + \HH_\text{int}$. 
 The Bogoliubov transformation also modifies the magnon dissipators
\begin{eqnarray}\label{eq:L_SWT}
\mathcal{L}_{\mop_j\mop^\dagger_j}[\hat \rho_{\rm mc}] =&& -s_\alpha c_\alpha\left(\mathcal{L}_{\mop_{\text{B},j}\mop_{\text{B},j}}[\hat \rho_{\rm mc}] + \mathcal{L}_{\mop^\dagger_{\text{B},j}\mop^\dagger_{\text{B},j}}\right) + \nonumber\\
    &&+ c_\alpha^2 \mathcal{L}_{\mop_{\text{B},j}\mop^\dagger_{\text{B},j}} + s_\alpha^2\mathcal{L}_{\mop^\dagger_{\text{B},j}\mop_{\text{B},j}}.
\end{eqnarray}
We choose $\tanh 2\alpha = 2\sq/\md$ to cast the total Hamiltonian $\HH_\text{tot}$  into the sum of a diagonal part, $\HH_\text{diag} = \sum_{j=1,2}\hbar \Bd \mop_{\text{B},j}^\dagger\mop_{\text{B},j} + \hbar\cd\hat c^\dagger \hat c$ and an off-diagonal interaction part
\begin{eqnarray}
    \HH_\text{int}/\hbar = \!\sum_{j=1,2}\!\gmc\hat c^\dagger\left(  c_\alpha \mop_{\text{B},j} - s_\alpha \mop_{\text{B},j}^\dagger\right) + \text{H.c.},
\end{eqnarray}
which can be treated perturbatively by the Schrieffer-Wolff transformation (SWT) generated by
\begin{eqnarray}
    \hat S = \!\sum_{j=1,2}\!\hat c\left(  \frac{\gmc c_\alpha}{\cd - \Bd} \mop_{\text{B},j} - \frac{\gmc s_\alpha}{\cd + \Bd} \mop_{\text{B},j}^\dagger\right)\!-\!\text{H.c.}
\end{eqnarray}
The transformed Hamiltonian, $e^{\hat S}\HH_\text{tot}e^{-\hat S}$, can be truncated to second order in the perturbative parameter $\left(\gmc/\left(\cd\mp\Bd\right)\right)\left(\left(\md \pm \Bd\right)/\left(2\Bd\right)\right)^{1/2}$, yielding $\HH_\text{eff} = \HH_\text{diag} + [\hat S, \HH_\text{int}]/2$~\cite{bravyi2011schrieffer}. Dropping the uncoupled cavity terms and reversing the Bogoliubov transformation, $\mop_j = c_\alpha \mop_{\text{B},j} - s_\alpha\mop_{\text{B},j}^\dagger$, we obtain Eq.~\eqref{eq:Heff}.

The magnon operators transform as
\begin{multline}\label{eq:sB_SWT}
    e^{\hat{S}}\mop_{\text{B},1}e^{-\hat{S}} = \\ =\beta_1 \hat{c} + \beta_2 \hat{c}^\dagger + \beta_3\mop_1 + \beta_4\mop_1^\dagger + \beta_5\mop_2 + \beta_6\mop_2^\dagger,
\end{multline}
up to second order in the perturbative parameter where $\beta_1 \equiv \gmc s_\alpha/\left(\cd + \Bd\right)$, $\beta_2 \equiv \gmc c_\alpha/\left(\cd - \Bd\right)$, $\beta_5\equiv \left(\beta_2^2 - \beta_1^2\right) c_\alpha/2$, $\beta_6 \equiv \left(\beta_2^2 - \beta_1^2\right) s_\alpha/2$, $\beta_3 \equiv c_\alpha + \beta_5$ and $\beta_4 \equiv s_\alpha + \beta_6$. The transformation for $\mop_{\text{B},2}$ follows from Eq.~\eqref{eq:sB_SWT}, by exchanging all magnon mode indices $1 \leftrightarrow 2$ on the right-hand side.
Similarly, for the cavity operators
\begin{equation}\label{eq:c_SWT}
        e^{\hat{S}}\hat ce^{-\hat{S}} = \mu_1 \hat{c} + \mu_2\left(\mop_1 + \mop_2\right) + \mu_3\left(\mop_1^\dagger + \mop_2^\dagger\right),
\end{equation}
where $\mu_1 \equiv 1 + \beta_2^2 - \beta_1^2$, $\mu_2 \equiv\beta_2 s_\alpha - \beta_1 c_\alpha$ and $\mu_3 \equiv \beta_2 c_\alpha - \beta_1 s_\alpha$. These expressions can be used to construct the transformed cavity and magnon dissipators, $e^{\hat S} \mathcal{L}_{\hat c\hat c^\dagger}[\hat \rho_{\rm mc}]e^{-\hat S}$ and $e^{\hat S} \mathcal{L}_{\mop_j\mop^\dagger_j}[\hat \rho_{\rm mc}]e^{-\hat S}$, respectively.

After the SWT and reversing the Bogoliubov transformation, the master equation contains dissipators of the form $\mathcal{L}_{\hat{v}_j\hat{v}_k}[\hat \rho_\text{mc}']$, where $\hat \rho_\text{mc}'$ is the final transformed density matrix and $\mathbf{\hat v}\equiv (\hat s_1,\hat s_1^\dagger, \hat s_2, \hat s_2^\dagger)^T$, as well as $\mathcal{L}_{\hat{v}_j\hat c_k}[\hat \rho_\text{mc}']$ with $\mathbf{\hat c} \equiv (\hat c, \hat c^\dagger)^T$ and their Hermitian conjugates, and $\mathcal{L}_{\hat c_j \hat c_k }[\hat \rho_\text{mc}']$.
The dissipators $\mathcal{L}_{\hat{v}_j\hat c_k}[\hat \rho_\text{mc}']$ oscillate in the interaction picture at frequencies $\vert\Omega_{\pm}\vert=\vert\left(\left(\mdSWT + \gSWT\right)^2 - \left(2\sqSWT + \GSWT\right)^2\right)^{1/2} \pm \left(\cd + 2\gmc\mu_3\right)\vert$,
where "$+$" corresponds to $\mathcal{L}_{\mop_j\hat c}$ and $\mathcal{L}_{\mop_j^\dagger\hat c^\dagger}$ and "$-$" corresponds to $\mathcal{L}_{\mop_j^\dagger\hat c}$ and $\mathcal{L}_{\mop_j\hat c^\dagger}$. The rates associated with these dissipators in the Schr\"odinger picture are $\Gamma_{\mop_j\hat c} = \gamma_\text{m}\mu_1\mu_3$, $\Gamma_{\mop_j^\dagger\hat c} = 0$, $\Gamma_{\mop_{j}\hat c^\dagger} = (\kappa-\gamma_\text{m})\mu_1\mu_2$ and $\Gamma_{\mop_{j}^\dagger\hat c^\dagger} = \kappa\mu_1\mu_3$.
Provided that $\vert\left(2\sqSWT + \GSWT\right)/\left(\mdSWT+\gSWT\right)\vert\lesssim 1$ and that $\vert\Omega_{+}\vert \gg \max\{\vert\Gamma_{\mop_{j}\hat c}\vert, \vert\Gamma_{\mop_{j}^\dagger\hat c^\dagger}\vert\}$ and $\vert\Omega_{-}\vert \gg \vert\Gamma_{\mop_{j}\hat c^\dagger}\vert$,
the magnon-cavity dissipators $\mathcal{L}_{\hat{v}_j\hat c_k}[\hat \rho_\text{mc}']$ and their Hermitian conjugates can be neglected under the rotating wave approximation. The two magnons and the cavity become fully decoupled, justifying the replacement of $\hat \rho_\text{mc}'$ by $\hat \rho_\text{m}$ in the main text.

The remaining rates in the two-magnon master equation (as defined in the main text) are
\begin{subequations}
\begin{eqnarray}
    &&\Gamma_{21} = \Gamma_{43} = \Gamma_{41} = \Gamma_{23} = \kappa \mu_3^2 \\
    &&\Gamma_{12} = \Gamma_{34} = \kappa \mu_2^2 + \frac{1}{2}\gamma_\text{m}\left(\mu_1^2 +1\right)\\
    &&\Gamma_{14} = \Gamma_{32} = \kappa \mu_2^2 + \frac{1}{2}\gamma_\text{m} \left(\mu_1^2 -1\right)\\
    &&\Gamma_{11} = \Gamma_{22} = \Gamma_{13} = \Gamma_{31} = \Gamma_{24} = \Gamma_{42} = \kappa \mu_2\mu_3\label{eq:Gamma_sq}.
\end{eqnarray}
\end{subequations}

\section{Analytical approximation for $\nu$}

To quantify entanglement, we construct the covariance matrix $\mathbf{\bar \sigma} = \left(\langle\mathbf{\hat q}\mathbf{\hat q}^T\rangle + \langle\mathbf{\hat q}\mathbf{\hat q}^T\rangle^T\right)/2$ where $\mathbf{\hat q} \equiv \left(\hat q_1, \hat p_1, \hat q_2, \hat p_2\right)^T$ is related to the ladder operators via $\mathbf{\hat q} = \mathbf{T}\mathbf{\hat v}$. 
The covariance matrix $\mathbf{\bar \sigma}$ is a $4\times4$ matrix, which can be subdivided into $2\times2$ matrices, $\mathbf{\bar \sigma} = \left(\begin{smallmatrix}
    \mathbf{A} & \mathbf{C}\\
    \mathbf{C}^T & \mathbf{B}
\end{smallmatrix}\right)$. The Duan-Simon criterion uses the smallest symplectic eigenvalue~\cite{duan2000inseparability, simon2000peres} given by
\begin{equation}\label{eq:nu}
    	\nu = \sqrt{\frac{1}{2}\left( \Delta - \sqrt{\Delta^2 - 4\det \mathbf{\bar\sigma}}\right)},
\end{equation}
where $\Delta \equiv \det A + \det B - 2\det C$ and $\nu<1/2$ indicates entanglement. 

Since the full covariance matrix $\mathbf{\bar\sigma}$ is dense, we consider a simplified model neglecting $\GSWT$ (valid for $\vert \GSWT\vert \ll 2\vert\sqSWT\vert$) and the dissipation rates from Eq.~\eqref{eq:Gamma_sq} (valid for $\vert\Gamma_{11}\vert\ll\vert\mdSWT + \gSWT\vert$ and $\vert\Gamma_{11}c_{(2\alpha_-)}\vert^2\ll(\mdSWT-\gSWT)^2-4\sqSWT^2$ and $\vert 4\Gamma_{11}\left(\begin{smallmatrix}
    \tanh_{\alpha_-}\\\coth_{\alpha_-}
\end{smallmatrix}\right)/\left(\Gamma_{12}-\Gamma_{14}\right)\vert\ll1$).
Defining $\mop_\pm = (\mop_1 \pm \mop_2)/\sqrt{2}$ and applying to the simplified master equation a Bogoliubov transformation $\mophB_\pm = c_{\alpha_\pm}\mop_\pm + e^{-i\varphi_\pm}s_{\alpha_\pm}\mop_\pm^\dagger$ where $\tanh\alpha_\pm = 2\sqSWT/(\mdSWT \pm \gSWT)$ and $\varphi_\pm=0$, yields the transformed master equation
\begin{eqnarray}
	\frac{\dd \hat \rho}{\dd t} =&& -\frac{i}{\hbar}\left[\HH_\text{B}, \hat \rho\right]
	+ \sum_{\chi=\pm}\sum_{jk}\Gamma_{\chi jk}\mathcal{L}_{\hat v_{\chi j}\hat v_{\chi k}}[\hat \rho]\label{eq:master_trans},
\end{eqnarray}
where $\HH_\text{B}/\hbar \equiv \Delta_+\mophB_+^\dagger\mophB_+ + \Delta_-\mophB_-^\dagger\mophB_-$ with $\Delta_\pm = \sqrt{\left(\mdSWT \pm \gSWT\right)^2 - 4\sqSWT^2}$ and $\mathbf{\hat v}_\chi = (\mophB_\chi, \mophB_\chi^\dagger)^T$. We neglect the term $\Gamma_{+ 11}\mathcal{L}_{\mophB_+\mophB_+}[\hat \rho]$ and its Hermitian conjugate under the rotating wave approximation, $\vert\Gamma_{+11}\vert \ll \vert2\Delta_+\vert$. The modified dissipation rates are given by
$\Gamma_{\chi 12} =\left(\Gamma_{12}+\chi\Gamma_{14}\right)c_{\alpha_\chi}^2 + 2\Gamma_{21}s_{\alpha_\chi}^2\delta_{\chi+}$, $\Gamma_{\chi 21} = \left(\Gamma_{12}+\chi\Gamma_{14}\right)s_{\alpha_\chi}^2 + 2\Gamma_{21}c_{\alpha_\chi}^2\delta_{\chi+}$ and $\Gamma_{\chi 11} = \Gamma_{\chi 22} = -\left(\Gamma_{12} +\chi \Gamma_{14}\right)s_{\alpha_\chi}c_{\alpha_\chi} - 2\Gamma_{21}s_{\alpha_\chi} c_{\alpha_\chi}\delta_{\chi+}$. At $\vert\mdSWT\vert \approx \vert g_{12}\vert$, $\Delta_+$ becomes imaginary, signaling the parametric instability threshold of the \textit{collective mode} $\hat{s}_+$ (thin instability region in Fig.~\ref{fig:3}). 

Eq.~\eqref{eq:master_trans} gives exact steady-state expectation values
\begin{subequations}\label{eq:expectations}
\begin{eqnarray}
    \langle\mophB_\chi^\dagger\mophB_\chi\rangle_{ss} =&& \frac{\Gamma_{\chi 21}}{\Gamma_{\chi 12} - \Gamma_{\chi 21}} \equiv n_\chi\\
	\langle\mophB_+^2\rangle_{ss} =&& 0\\
	\langle\mophB_{-}^2\rangle_{ss} =&& \frac{\Gamma_{-11}}{\Gamma_{-21} - \Gamma_{-12} - 2 i \Delta_-} \equiv z,
\end{eqnarray}
\end{subequations}
which are then used to reconstruct $\mathbf{\bar{\sigma}}$ via inverse transformations. Setting $n_+, \alpha_+ \simeq 0$ (valid for $2\vert \Gamma_{21}\vert\ll \vert \Gamma_{12}+\Gamma_{14}\vert$ and $2\vert \sqSWT\vert\ll\vert\mdSWT+\gSWT\vert$) recovers Eq.~\eqref{eq:nu_0}.

Including uncorrelated thermal noise modifies the covariance matrix as $\mathbf{\bar\sigma} \to \eta \mathbf{\bar\sigma} + (1-\eta)\mathbb{1}$, reproducing Eq.~\eqref{eq:nu_eta}.

\end{document}